\documentclass[letterpaper,twocolumn,10pt]{article}
\pdfoutput=1
\usepackage{styles/usenix-2020-09}

\usepackage[sort&compress,numbers]{natbib}
\usepackage{amsmath,amssymb,amsfonts}
\usepackage{graphicx, subcaption}
\usepackage{textcomp}
\usepackage{xcolor}
\usepackage{balance}
\usepackage{comment}
\usepackage{multirow}
\usepackage{siunitx}    
\usepackage{soul}
\usepackage{textcomp}
\usepackage{tabularx}
\usepackage{algorithm}
\usepackage{algpseudocode}
\usepackage{xspace}
\usepackage[draft]{todonotes}
\usepackage{listings}
\usepackage{styles/c_custom_style}
\usepackage{xparse}
\usepackage{courier}

\newcommand{\shortname}{\textsc{OCA}\xspace}
\newcommand{\cmk}{\checkmark}

\begin{document}

\date{} 

\title{On-the-fly Code Activation for Attack Surface Reduction}
\author{
{\rm Chris Porter} \\
Georgia Institute of Technology
\and
{\rm Sharjeel Khan} \\
Georgia Institute of Technology
\and
{\rm Santosh Pande} \\
Georgia Institute of Technology
}

\maketitle



\begin{abstract}

Modern code reuse attacks are taking full advantage
of bloated software. Attackers piece together
short sequences of instructions in otherwise benign code
to carry out malicious actions.
Eliminating these reusable code snippets, known as gadgets,
has become one of the prime concerns of attack surface
reduction. The aim is to break these chains of gadgets,
thereby making
such code reuse attacks impossible or substantially
less common.
Previous work on attack surface reduction has typically
tried to eliminate such attacks by subsetting
the application, e.g. via user-specified inputs, configurations,
or features, or by focusing on third-party libraries
to achieve high gadget reductions with minimal interference
to the application.

In this work we present a general, whole-program attack
surface reduction technique called \shortname that significantly
reduces gadgets and has minor performance
degradation. \shortname requires no user inputs and leaves all
features intact. \shortname identifies specific program points
and through analysis determines key function sets to enable/disable
at runtime. The runtime system, thus, controls the set of
enabled functions during execution, thereby significantly
reducing the set of active gadgets an attacker can use, and
by extension, cutting down the set of active gadget
chains dramatically.
On SPEC CPU 2017, our framework achieves 73.2\% total gadget
reduction with only 4\% average slowdown. On 10 GNU coreutils
applications, it achieves 87.2\% reduction.
On the nginx server it achieves 80.3\% reduction with 2\% slowdown.
We also provide a gadget chain-breaking study across all applications,
and show that our framework breaks the shell-spawning chain in all cases.



\end{abstract}

\section{Introduction} \label{sec:introduction}

Attack surface reduction has gained in importance lately.
A recent investigation
\cite{piece-wise} showed that on average, 95\% of GNU libc code \cite{glibc}
is never used by user applications in a typical
Ubuntu Desktop installation.
This bloated software has security implications.
The excess code can contain bugs and is typically not maintained well.
This allows it to become a landmine of vulnerabilities, or
to be repurposed for malicious ends in a code reuse attack.


In this work we aim to reduce the exposed, reusable
portions of code that are available when an attacker
launches a code reuse attack.
We motivate our work with a simple but powerful example.
It shows why attack surface
reduction is important, and also why it is challenging
to make a meaningful dent in the reusable code used for staging the attack.

\begin{lstlisting}[
  caption={An ROP gadget chain from \texttt{nginx} that spawns a shell with \texttt{execve}.},
  label=lst:rop-chain-execve,
  language=C,
  style=c,
  numbers=none,
  stringstyle=\color{black}
]
gadget 1 addr  -> pop rax; ret;
'/bin/sh'
gadget 2 addr  -> pop rcx; ret;
bss addr       -> 0x00800000
gadget 3 addr  -> mov qword ptr [rcx], rax; ret;
gadget 4 addr  -> pop rax; ret;
0x00000000
gadget 5 addr  -> pop rcx; ret;
bss addr + 8   -> 0x00800008
gadget 6 addr  -> mov qword ptr [rcx], rax; ret;
gadget 7 addr  -> pop rdi; ret;
bss addr       -> 0x00800000
gadget 8 addr  -> pop rsi; ret;
bss addr + 8   -> 0x00800008
gadget 9 addr  -> pop rdx; ret;
bss addr + 8   -> 0x00800008
gadget 10 addr -> pop rax; ret;
0x0000003b
gadget 11 addr -> syscall;
\end{lstlisting}

Listing \ref{lst:rop-chain-execve} shows an example of
an ``ROP gadget chain.'' These chains
can be used as part of an attack to
leak secrets, hijack processes, or otherwise cause
damage. We provide more background on
return-oriented programming (ROP) and gadgets
in the next section, but for the unfamiliar,
the high level idea is this: Snippets of code (gadgets)
are strung together with return statements to jump
from one to the next, in order to carry out some malicious
computation to launch an attack.

In the above example, the chain is designed to compute relatively little
but to a powerful end: It will launch a shell
via \texttt{execve}. This is a real chain that
has been automatically 
generated by the Ropper tool \cite{ropper} on the
\texttt{nginx} server application \cite{nginx}. Launching such an attack
requires exploiting a memory vulnerability, which is
also a realistic possibility here.
CVE-2013-2028, for example, is a bug in the decoding functionality
of \texttt{nginx}. Carlini et al \cite{control_flow_bending} show how to exploit this bug to write to arbitrary locations.
Note that as with any code reuse attack that we consider in this work,
we assume the attacker has some way of initiating the attack.
Here we assume, for example, that the attacker can
exploit some memory vulnerability to write this chain into
the stack, overwrite the return address to point to
gadget 1's address, and freely return along the ROP chain. Due to C's weak memory model and high costs of full memory protection, such vulnerabilities unfortunately do exist within almost all software written in C. 

Referring to Listing \ref{lst:rop-chain-execve},
the ROP gadget chain proceeds as follows. It stores the address
of the ``/bin/sh'' string into some location in the .bss section
(gadgets 1-3). Then, 8 bytes beyond that, it stores the value 0
(gadgets 4-6). Then it places the address of ``/bin/sh''
into the \texttt{rdi} register, and it places the 0 value into \texttt{rsi} and
\texttt{rdx} (gadgets 7-9). Lastly, it stores the \texttt{execve}
syscall number 0x3b
into the \texttt{rax} register and then executes the syscall instruction
(gadgets 10-11). 

The actual control flow for executing this attack is return-based.
The gadget addresses are the return targets. Thus,
each \texttt{ret} instruction
does the attacker's bidding, popping the gadget address and
jumping to the next malicious snippet.

There are several important details about the strength
of gadget chains.
First, as in this case, the chain does not need to be Turing-complete
to be effective.
Here the gadgets perform a
specific set of actions, all in the service of setting
up an \texttt{execve} call that will launch \texttt{/bin/sh}.
Second, the gadget chain does not need to be built
from ``intended'' instructions on architectures such
as x86. Though not shown in the listing, gadget
addresses from this \texttt{nginx} example leverage
``unintended'' instructions.
They are at unintended offsets in the original
instructions (which is allowed in x86's variable-length
instruction set architecture).
Third, the gadgets are not particularly rare instruction
sequences, so scavenging these gadgets (or computationally
equivalent sequences) from the executable code ought
to be achievable given a sufficient set of instructions.

These points illustrate the importance as well as the difficulty of the problem.
Not only is it possible to create an attack from only
a handful of gadgets from a large pool of instructions,
but on x86 the gadgets can be
cherrypicked from what is essentially a byte stream
of executable code.
And this attack represents a very serious security breach. Once the attacker
hijacks the process and launches a shell, they
can perform any action with that process' permissions.

Defending against code reuse attacks is still
an open research problem. Though we have made several
assumptions in this simple example, more
complex gadget-based attacks exist.
Current defenses are not able to handle them.
A critical question is whether the attack surface
can be reduced substantially enough to break such chains.

Thus, we propose \shortname, an attack surface
reduction technique for reducing reusable gadgets and
breaking their chains.
This paper makes the following contributions (please refer to
Section ~\ref{sec:evaluation} for results and their analysis):
\begin{enumerate}
    \item The first sound, whole-application technique for on-demand loading
          and purging for attack surface reduction which also has low runtime overhead.
    \item An evaluation on SPEC CPU 2017, GNU coreutils, and \texttt{nginx} that is
          comparable to unsound techniques in terms of gadget reduction.
    \item Evidence that short but highly detrimental gadget chains can be broken
          by this technique.  
\end{enumerate}

In the next section we present details of the problem
and closely related solutions.
Then we give an overview of our solution in Section \ref{sec:overview},
including assumptions.
Then we provide details of the framework in Section \ref{sec:framework},
followed by an evaluation in Section \ref{sec:evaluation}.
Lastly we provide more related work (Section \ref{sec:related-work})
and conclude (Section \ref{sec:conclusion}).

\section{Background and Motivation}\label{sec:background}


Code injection could be considered the predecessor of
modern code reuse attacks.
Early code injection attacks could simply inject code into memory
such as the heap and execute it. This was countered by
data execution prevention (DEP) \cite{dep}.
DEP enforces the write XOR execute (W $\oplus$ X) property on pages,
which is sufficient for stopping such blatant attacks.

Attackers grew to overcome this.
Perhaps the most basic code reuse attack is the classic
return-to-libc attack \cite{Nergal, expressiveness_retlibc}.
In a return-to-libc attack, the attacker
exploits some memory vulnerability to compromise ``control data,''
i.e. an indirect branch instruction
(return, indirect jump, or indirect call).
Then the attacker redirects control flow to jump into
GNU libc (glibc) code, which is full of functionality
that can be very useful to  an attacker.
For example, one useful place to jump can be
the \texttt{mprotect} function.
If the attacker can control the parameters to this function, then they
can remap a page as writable, breaking the  W $\oplus$ X property,
and allowing them to carry out code injection.
Address space layout randomization (ASLR) \cite{aslr} can make
it more difficult to locate target code such as the glibc library
calls, but then there are also known attacks that get around it
\cite{bypassing_aslr, jump_over_aslr, effectiveness_aslr, jit_aslr}.

Reuse attacks have only become more complicated. 
Return-oriented programming \cite{rop}, jump-oriented programming \cite{jop, jop2},
and call-oriented programming \cite{pcop} are all techniques that
leverage existing code to perform attacks. They rely on ``gadgets'' in the
code base, which are sequences of instructions that can be strung
together to perform some computation. In fact, an attacker
can construct a
Turing-complete program from gadgets that exist in the code base,
but this is often not even needed.
Turing-incomplete functionality may be sufficient if it still
allows the attacker to carry out some useful action (e.g. leaking a
secret, which could be the end goal, or spawning a shell, to give the
attacker more control).


Traditional defenses to these attacks have had some success,
but they have also had their shortcomings
(see Section \ref{sec:related-work}).
Due to these and other reasons, attack surface reduction is one class of defense
that has gained in prominence lately.
Piece-wise compiler \cite{piece-wise}, 
Chisel \cite{chisel}, Razor \cite{razor}, and BlankIt \cite{blankit}
are four recently developed debloating/attack surface reduction techniques that motivate 
our work.
They successfully show how to reduce
applications' attack surfaces, but they have shortcomings,
as well.

Piece-wise compiler \cite{piece-wise} modifies the loading stage at process start-up
to remove unreachable library code. It is sound, removing only
unneeded functionality from libraries (for which it requires
libraries' source code). No user input is needed, but piece-wise-compiled
libraries must be provided to a program before running it.
This approach removes function(s) in the library only if they
are proved unreachable on a whole-application basis,
i.e. from nowhere in the driving application can they ever be
invoked. Due to complex control flow in the libraries and
conservative limitations of static analysis, proving such
a property becomes extremely difficult. As a result this
work has not been demonstrated on glibc, a real world library
with vulnerabilities; on musl-libc its success is limited, leading
to debloating roughly 73\% of its gadgets when linked with SPEC CPU 2006. 

Chisel uses reinforcement learning to learn
which parts of a program are actually used and needed, and then builds
a trimmed version of it. Chisel is an unsound technique. It may
learn a model that eliminates needed functionality. This
can induce crashes and so is not practical in the general case.
Chisel works in a kind of compile-test-refine loop, as its
learner identifies which parts of the program are needed
in order not to crash or provide bad output. As such, it requires
a user specification (supplied as test inputs) and source code.
Chisel is designed to work with application code.

Razor uses heuristics and test inputs to debloat binaries.
Like Chisel, Razor is unsound and can lead to crashes.
Razor is designed for and works on applications, though they
include a discussion on its current effectiveness on libraries.
Similar to Chisel and Piece-wise, it also cannot debloat
``may-use code.'' That is, when these techniques remove a
piece of code, it is because that code is presumed unnecessary.

BlankIt is a binary runtime technique for
libraries that loads the library code that is needed
on each call site in a dynamic sense, i.e. on each call
to a library from the driving application.
BlankIt achieves a very high precision of attack
surface reduction of 97.5\% and removes all known CVEs
from real world libraries such as glibc. It is able to
achieve this precision due to call-specific predictions
of reachability that stem from the ML-based analysis of
its arguments. It is able to thus carry out debloating
of ``may-use code'' due to runtime analysis. It accomplishes
attack surface reduction by
using a binary runtime framework to load and unload library code
as needed. The runtime also has a predictive component that
can flag unexpected control flow.
BlankIt is however limited only to libraries and does not work on application code.

We summarize and compare the  characteristics of these four approaches in
Table \ref{table:debloat-comparison}.
With regard to practical usage, we find the last two rows strikingly important
(i.e. whether it is sound, and whether it can debloat may-use code).
Regarding soundness, this is critical for a debloating
technique to achieve general
adoption. It cannot make unsound transformations that can
crash applications.
Regarding whether the technique can debloat may-use code,
this is critical for diminishing the ``blast radius''
of an attack.
\emph{Solutions
such as Razor, Chisel, and Piece-wise do not reduce the attack surface
for any code that may be needed}.

\begin{table}[h!]
 \centering
  \caption{Broad properties of 4 state-of-the-art debloating techniques for security
  (Piece-wise, Chisel, Razor, and BlankIt).}
 \begin{tabular}{|l|l|l|l|l|}
   \hline
    & \bf PW & \bf Chsl  &  \bf Rzr  &  \bf BI \\
   \hline
   \hline
    Works on application    &     & \cmk    & \cmk &  \\
    Works on library	    & \cmk    &     & \cmk & \cmk   \\
    Works on binary	        &     &     & \cmk & \cmk \\
    No user input needed	& \cmk    &     &     & \cmk    \\
    No training needed	    & \cmk     &     & \cmk    &     \\
    Is sound	            & \cmk    &     &     & \cmk     \\
    Can debloat may-use code	&     &     &     & \cmk    \\
   \hline
  \end{tabular}
  \label{table:debloat-comparison}
\end{table}

Some of the above problems stem from the underlying techniques used,
especially involving machine learning. 
One of ML's typical drawbacks is
identifying realistic and plentiful training data.
This problem is present in both BlankIt and Chisel.
Another drawback can be the negative impact of mispredictions.
In Chisel, the prediction is at compile-time, and misprediction leads
to soundness issues.
In BlankIt, prediction is at runtime, which has two issues:
(1) how to act when an alarm is raised (which could be a false positive);
and (2) how to ascertain whether the input data to the predictor is trustworthy.
The inputs to predictive models
are susceptible to memory corruption vulnerabilities, just as other parts of the
program are. BlankIt attempts to alleviate the problem by hoisting the values and their predictions as high up in the control flow as possible which can potentially help
with this, but it is still not a guarantee.

Both Chisel and Razor require a user specification of
what functionality an application needs.
In both cases this specification is provided as test cases.
Not only can this requirement limit adoption for real-world use, in both of
these works it is tied directly to the soundness. If
the user does not provide the ``right'' set of test cases,
then the debloated program may crash. In their evaluation, several such cases are shown which is a significant issue.

To the best of our knowledge, there is no general technique today
that (1) works on the applications as a whole instead of libraries,
(2) is sound, and (3) can debloat may-use code.
Current techniques either tackle libraries
to achieve strong attack surface reduction, or they
tackle applications and compromise soundness.
Furthermore, an ideal solution would not require any test cases
or specification from the user; and it would either avoid
prediction or handle its security challenges gracefully.
All these limitations motivate the current work.

\section{Overview} \label{sec:overview}

\subsection{Proposed solution}

We propose a compiler and runtime solution for attack
surface reduction called \shortname. It is sound, has a static and runtime component,
requires no user input, requires no hardware changes, and works
on application code.

\shortname embraces the idea that only code that is currently needed
by a running program should be available for execution;
the rest should be made inaccessible such that any attempt
to access it should trigger a runtime exception.
Active sections of code form ``decks'' that the program can effectively
stand on. When a deck is unneeded, it can be removed. To
take the analogy further, \shortname is a technology for
attack surface reduction,
but it can be viewed as \textit{constructive}.
It achieves attack surface reduction not by cutting down the
program, but by putting together
the active code that it needs at any particular execution point.
A deck could be a group of functions which are guaranteed to be
executed from current execution point. Since such a set cannot
be precisely generated without causing heavy runtime overheads
(especially inside loops), \shortname will turn this into a
tight overestimation problem inside respective regions.

\shortname depends heavily on static analysis. The decks of the program 
are based off of static features. An outermost loop in a loop nest must
be treated as its own deck, for example.
Similarly, a single-function leaf node in
the callgraph, which is not reachable by any loop, forms its own
simple deck.
In our implementation, granularity of disabling/enabling mechanism is
at the system page level.
Creating and tearing down a deck corresponds
to marking code pages read-execute (RX) and read-only (RO),
respectively. In other words, if a deck consisting of functions
foo() and boo() is to be turned on or enabled at some program point,
one must execute calls to mark the respective pages that contain
 foo() and boo()'s code as read-execute (RX).  

\begin{figure}[h]
    \centering
    \includegraphics[scale=0.3]{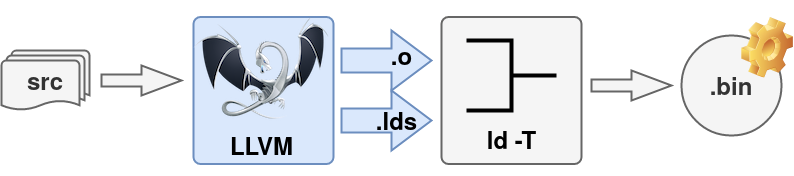}
    \caption{High-level view of the compiler part. The LLVM pass outputs a custom linker script and instrumented object file.}
    \label{figure:compiler}
\end{figure}

Figure \ref{figure:compiler} shows a high-level view of
the compiler step of the \shortname solution. 
Application source code is fed into the LLVM compiler pass.
The compiler
performs static analysis and identifies programs points for decks to be
enabled (RX) and disabled (RO); it also performs partitioning
of such sets of functions to improve security benefits as discussed later.
Additionally, the pass creates a custom linker script. Both of
these are fed into the linker (where the \texttt{-T} options consumes
the custom linker script). The linker produces the final binary.
\shortname is represented by the blue part in the figure,
and the linker itself is unmodified.

\begin{figure}[h]
    \centering
    \includegraphics[scale=0.4]{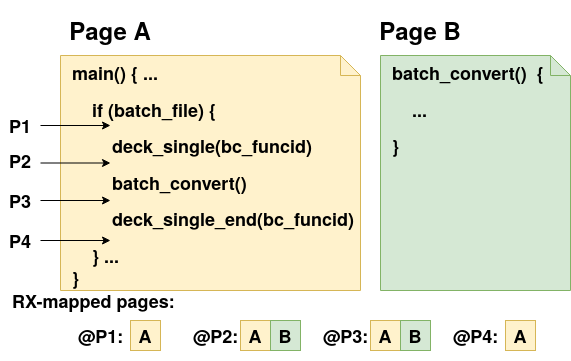}
    \caption{High-level runtime example from GNU coreutils' \texttt{date}.
    The set of RX-mapped pages increases at P2 to include a called function.}
    \label{figure:runtime}
\end{figure}

Figure \ref{figure:runtime} illustrates the runtime by way of an example.
It is drawn directly from the GNU coreutils' \texttt{date} program \cite{coreutils},
which provides a command-line option for reading dates from
a file (given by the \texttt{-f} switch). When this option
is given, \texttt{main} invokes \texttt{batch\_convert}.
At runtime, \shortname will create and tear down
a deck for this single function, \texttt{batch\_convert}.
In Figure \ref{figure:runtime}, there are two code pages in memory
(for simplicity).
Page A contains \texttt{main}, and page B contains \texttt{batch\_convert}.
The call to \texttt{batch\_convert} has been instrumented by the compiler,
so that before it is invoked, its page will be mapped RX, and after
it returns, its page will be mapped RO. These mapping steps are done
by \texttt{deck\_single} and \texttt{deck\_single\_end}, respectively.
\texttt{bc\_funcid} is the function ID for \texttt{batch\_convert}
assigned by \shortname at compile-time.
The 4 program points, P1-P4, indicate which pages are
mapped RX at each step.
One can see that gadgets in unmapped decks and their pages
are thus inaccessible to the attacker;
thus, the finer the granularity of the deck
(finest granularity being one function), the better the security.

\subsection{Threat Model}
We assume the operating system and compiler are trusted.
The source code and any
third-party libraries may contain bugs.
For simplicity, we do not handle dynamically generated code or
self-modifying programs; we focus on C/++ binaries.

We are focused solely on attack surface reduction
and assume the attacker has some way of initiating and
propagating the attack (e.g. that the attacker can exploit
a memory vulnerability and trigger a gadget chain). Given
today's state-of-the-art defenses, we find this assumption
reasonable.

We assume the runtime is protected (a similar assumption in \cite{blankit}),
and which can be implemented with in-process isolation \cite{cpi}, hardware
segmentation or software fault isolation \cite{getting-point}.
This prevents attackers from jumping into the runtime and guarantees,
along with the trusted loader, that the statically computed
function IDs and framework metadata are protected.

Arguments to the runtime API are statically evaluated and passed by register
and so cannot be tampered with, except in the case of indirect calls.
How to guarantee the integrity of indirect call targets is
precisely the problem handled by orthogonal schemes like CFI and CPI
(see Section \ref{sec:related-work}).
\shortname does \textit{not} tackle this problem, which we
consider out of scope. \shortname is focused on reducing the
attack surface available to an attacker when an attack occurs,
but schemes such as CPI would still be needed for pointer integrity.
Similarly, redirecting control flow to another instrumented
runtime call that is mapped RX depends on orthogonal defenses.
Repeatedly invoking the same instrumented runtime call
that is mapped RX, however, is disallowed by construction (see
Section \ref{sec:framework}, which details how instrumented
calls will only execute exactly once for every paired teardown
call).

As described earlier, the threat is an attacker
exploiting the memory vulnerabilities
of an application executing under the \shortname system,
attempting to string together a gadget chain to launch an attack.
Due to needed gadgets residing in multiple decks that are disabled,
however, the attack will lead to a runtime exception and be caught.
\section{Framework} \label{sec:framework}
The framework consists of a compiler and runtime component.
The compiler is responsible for inserting calls in the application
to the runtime in order to create and tear down decks.
The runtime receives requests from the application
and enables and disables the code pages associated
with each deck.
In this section we discuss each part separately and then cover
optimizations.

\subsection{Compiler component} \label{compiler-component}
The compiler portion is an LLVM \cite{llvm} pass that
can be divided further into two parts:
instrumentation and linker script output.
During instrumentation, \shortname identifies function calls
and loops and instruments them appropriately with calls to the runtime.
As the pass does this, it collects critical static
information for organizing the text section, which
it uses to create a custom linker script.
The key idea of this work is to keep the
decks\footnote{A deck is defined as a group of functions that are enabled at a program point by turning their page permissions to RX}
as lean as possible. Ideally, each deck should consist of one
function, i.e. only one function should be enabled at a
time for highest security. Such a scheme would incur very high
overheads, however, especially for call chains that execute
inside loops, and would make the scheme untenable; thus, based on
the context surrounding a program point, static analysis identifies
what a deck should be and hoists the calls to the runtime accordingly. 

\subsubsection{Analyzing for Decks} \label{compiler-instrumentation}
Analysis and instrumentation for decks is heavily organized around loops.
Loops are problematic because adding code for enabling or disabling a deck inside
them can cause significant performance degradation.
We define two terms: \textbf{encompassed}
and \textbf{non-encompassed} functions to distinguish between the loop context that surrounds them.
A function is encompassed if it is called inside of a loop, or if it is
reachable within the callgraph by some function that is called within
a loop through a caller-callee relation. To determine the encompassed function set, the pass first identifies
all functions called within a loop, and then takes the transitive closure of
any functions reachable from that set using the caller-callee relation shown by the callgraph. The non-encompassed function set
is simply the set of all functions minus the encompassed function set.

\shortname's default treatment of loops is to bear on the side
of performance. Due to this reason, it tries to avoid instrumenting inside of them,
because if this function is called in a surrounding inter-procedural loop at runtime,
its deck's instrumentation will incur repeated invocations, leading to high overheads.
Interprocedurally this implies that it cannot instrument inside of
encompassed functions, either. This also raises a problem for
loop-enclosed indirect calls, whose static target set can be large,
and whose precise dynamic value is often unknown until execution is
inside of the loop.

To handle these different cases, the pass instruments
four different types of decks which will invoke the runtime:
(1) Single, (2) Loop, (3) Reachable, and (4) Indirect.
We describe each one, covering where it is placed,
and what it is responsible for mapping.

The \textbf{single} deck is used when a non-encompassed function
calls a non-encompassed function. This is the simplest case.
The compiler must ensure that the function being
called is mapped RX before it is actually executed. Because the callee
is known to be non-encompassed (i.e. not part of some transitive closure
that lies within a loop), only a single function needs to be mapped
RX (i.e. the callee itself).

The \textbf{loop} deck is placed at the outermost loop header
for any loop nest in any non-encompassed function.
It is designed to map all functions
that can be reached interprocedurally within that loop. Notice that
an encompassed function will never instrument a loop deck.
If it did, it could be invoked repeatedly within a possible interprocedural loop, which could
lead to severe slowdown. 

The \textbf{reachable} deck is placed in a non-encompassed
function before any calls to encompassed functions.
Even though encompassed functions are, by definition,
part of some interprocedural loop, they may also
be reachable in the callgraph via some non-loop path.
Encompassed functions should also not contain any instrumentation (due
to the potential performance degradation).
Thus, when a non-encompassed function invokes an encompassed one,
the compiler needs the \textbf{reachable} deck to
map the encompassed function and any interprocedurally
\textbf{reachable} functions from it as RX.

The \textbf{indirect} deck is for function pointers.
The challenge of function pointers is that their exact targets are
often not known statically, and therefore the compiler cannot determine
precisely what needs to be mapped RX until runtime. Function pointer
analysis can help narrow the possible targets but would still
be an overapproximated set at compile time
(which would limit attack surface reduction).
\shortname opts to solve this at runtime. The instrumentation
passes the function pointer to the runtime library, which
then maps the appropriate page(s).

Using function pointers' runtime values may require
\shortname to break its own rule of
disallowing instrumentation inside of loops since one of its targets could be an encompassed function. When
an indirect call is inside of a loop, \shortname
must still instrument it since its target may not be known outside the loop.
On the other hand, as mentioned, repeatedly executing instrumentation and
runtime library code inside of a loop can drastically
degrade performance. Such cases are handled via optimization (discussed in
Section \ref{optimizations}).

\begin{figure}[h]
    \centering
    \includegraphics[scale=0.20]{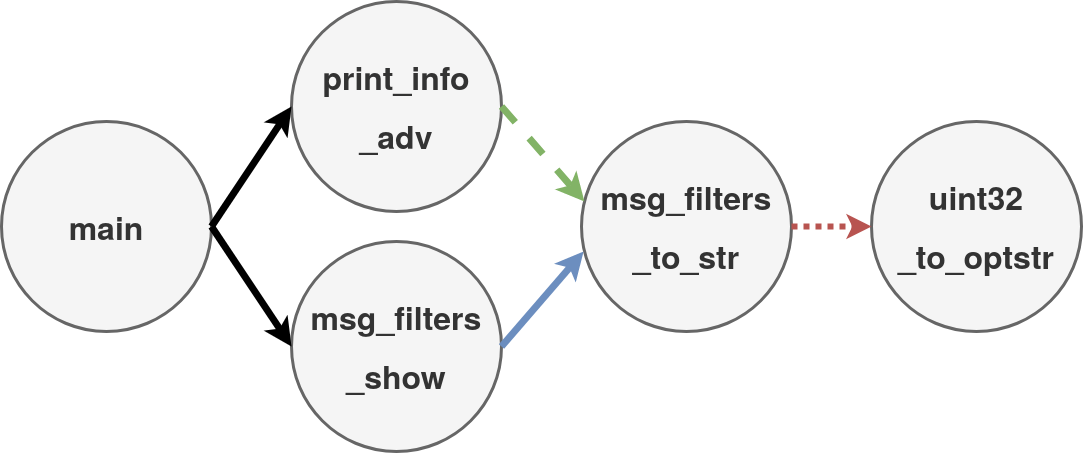}
    \caption{Simplified callgraph from the \texttt{xz} data compression
    application. This illustrates 4 types of edges,
    each of which requires different handling by
    the instrumentation pass.}
    \label{figure:cg-instrumentation}
\end{figure}

Figure \ref{figure:cg-instrumentation} depicts a sub-callgraph
from SPEC CPU 2017's \texttt{xz},
a data compression application \cite{spec2017}.
It illustrates all but the \textbf{indirect} case.
Each node in the figure is a function, and each
edge is a call. Only the call from \texttt{print\_info\_adv}
to \texttt{msg\_filters\_to\_str} (dashed, green)
is inside of a loop.
The set of encompassed functions is therefore
\{\texttt{msg\_filters\_to\_str}, \texttt{uint32\_to\_optstr}\},
and the set of non-encompassed functions is
\{\texttt{main}, \texttt{msg\_filters\_show}, \texttt{print\_info\_adv}\}.
Instrumentation will be treated as follows:
\begin{enumerate}
    \item The solid-black edges from \texttt{main}
        require \textbf{single} decks.
    \item The dashed-green edge from \texttt{print\_info\_adv}
        requires a \textbf{loop} deck;
        instrumentation will be at the loop preheader that dominates
        the call to \texttt{msg\_filters\_to\_str}.
    \item The solid-blue edge from \texttt{msg\_filters\_show}
        requires a \textbf{reachable} deck.
    \item The dotted-red edge from \texttt{msg\_filters\_to\_str} 
        will have no instrumentation, because it is encompassed.
\end{enumerate}

Basic pseudocode for the compiler instrumentation pass is shown
in Algorithm \ref{alg:compiler-pass}. It shows two functions,
\texttt{run\_on\_nonencompassed\_func} and \texttt{run\_on\_func},
which are hooks called by the pass manager on non-encompassed functions
and all functions, respectively.
The pseudocode shows the general logic for
how decks are selected and inserted.
Each deck needs only one key piece of runtime information, namely
a unique ID that is generated statically for each loop or function.
At runtime the library maps this parameter to a set of functions
and their corresponding pages in memory.
(In the case of indirect calls, the only difference is that the
runtime target address is used instead of a statically known ID.)
Several details not shown in the pseudocode
but that should be noted include: the insertion of 
the deck teardown calls; the insertion of an initialization
call at program start; construction of functions' static reachability;
and construction of the encompassed function set.

\begin{algorithm}[htb]
\caption{Pseudocode for \shortname's compiler pass.}
\label{alg:compiler-pass}
\begin{algorithmic}[]
\Function{run\_on\_nonencompassed\_func}{$func$}
    
    \For{$instr$ in $func$}
        \If{$instr.\Call{is\_loop\_body}$}
            \State \textbf{continue}
        \EndIf
        \If{$instr.\Call{is\_loop\_start}$}
            \State $\Call{insert\_deck}{LOOP, instr.id}$
        \EndIf
        \If{$instr.\Call{is\_direct\_call}$}
            \State $target \gets instr.\Call{get\_call\_target}$
            \If{$target.\Call{is\_encompassed}$}
                \State $\Call{insert\_deck}{REACHABLE, target.id}$
            \Else
                \State $\Call{insert\_deck}{SINGLE, target.id}$
            \EndIf
        \EndIf
        
    \EndFor
\EndFunction
\\
\Function{run\_on\_func}{$func$}
    \For{$instr$ in $func$}
        \If{$instr.\Call{is\_indirect\_call}$}
            \State $target\_addr \gets instr.\Call{get\_func\_ptr}$
            \State $\Call{insert\_deck}{INDIRECT, target\_addr}$
        \EndIf
    \EndFor
\EndFunction
\end{algorithmic}
\end{algorithm}

\shortname's compiler instrumentation supports non-trivial
C and C++ behavior. Though the details are unimportant, it
is important to stress that the approach is general. Some
of these features include the following:
In addition to handling LLVM IR's call instructions,
it must also handle invoke instructions and therefore landing pads.
It handles external libraries that take
and then invoke a callback to the application.
It handles recursion.
It handles C++ destructors that can be invoked when an exception is
thrown (via \texttt{\_\_cxa\_throw}).
It handles libc\_nonshared.a.
It handles start-up C++ code before \texttt{main}.
It handles signal handlers, including \texttt{atexit} and \texttt{on\_exit}.

\subsubsection{Linking} \label{compiler-linking}
At the end of the compiler pass, \shortname outputs a custom linker script.
Intuitively, the goal of the linker script is to separate
functions into different pages so that marking 1 function as RX does
not ``activate'' unrelated functions (i.e. make them and their gadgets
available for use).
Two examples are helpful for understanding this stage,
and we refer again to the \texttt{xz} callgraph example
in Figure \ref{figure:cg-instrumentation}.

\textbf{Example 1}:
Recall that \texttt{print\_info\_adv} and \texttt{msg\_filters\_show}
are single decks. Without any enforcement via the linker script,
these two functions could arbitrarily belong to the same
system page at runtime.
If \shortname allowed this,
then mapping one of these functions
RX could inadvertently map the other RX, which exposes
more code surface and therefore hurts overall security.
But because these are different decks,
the \shortname instrumentation guarantees
that each will be mapped RX independently
at runtime. Thus, the custom linker script can safely
separate these single decks to different page-aligned
sections. That is, the decks $S1 = \{print\_info\_adv\}$ and
$S2 = \{msg\_filters\_show\}$ will each
form their own disjoint set. Each disjoint set
is page-aligned by the custom linker script.

\textbf{Example 2}: We now consider the rest of the callgraph
in Figure \ref{figure:cg-instrumentation} and expand it by one node
(see the updated callgraph in Figure \ref{figure:cg-instrumentation-expanded} in
Appendix Section \ref{appendix:expanded-xz-cg}).
In \texttt{xz}, there is in fact a function named
\texttt{parse\_block\_header} between
\texttt{print\_info\_adv} and \texttt{msg\_filters\_to\_str}
in its callgraph.
It is on the same path and still part of the interprocedural
loop. Thus, the full callgraph's deck sets $Dk.*$ are as follows:
  \begin{equation*}
  \begin{aligned}
  & Dk.S1 = \{print\_info\_adv\} \\
  & Dk.S2 = \{msg\_filters\_show\} \\
  & Dk.L = \{parse\_block\_header, msg\_filters\_to\_str, \\
           & uint32\_to\_optstr\} \\
  & Dk.R = \{msg\_filters\_to\_str, uint32\_to\_optstr\}
 \end{aligned}
\end{equation*}

We follow a similar logic as in Example 1. Any of
these functions can arbitrarily belong to the same
system page at runtime.
Thus, without any enforcement via the linker script,
\texttt{print\_info\_adv} and \texttt{msg\_filters\_show}
could again reside in the same page. Similarly,
\texttt{parse\_block\_header} can occupy the same system
page as either of the functions in $Dk.R$. That is, invoking
\texttt{msg\_filters\_to\_str} from \texttt{msg\_filters\_show}
at runtime could
inadvertently activate the gadgets in \texttt{parse\_block\_header}.
The solution is to again rely on the fact that
\shortname instrumentation
ensures that each deck will be mapped RX independently at
runtime, and to leverage the custom linker script to
avoid this security penalty.
The custom linker script should separate the intersection
$L \cap R = \{msg\_filters\_to\_str, uint32\_to\_optstr\}$
into its own disjoint set. Thus, the full
disjoint sets $Dj.*$ are as follows:
  \begin{equation*}
  \begin{aligned}
  & Dj.S1 = \{print\_info\_adv\} \\
  & Dj.S2 = \{msg\_filters\_show\} \\
  & Dj.L.1 = \{parse\_block\_header\} \\
  & Dj.I.LR = \{msg\_filters\_to\_str, uint32\_to\_optstr\}
 \end{aligned}
\end{equation*}
---where
$Dj.I.LR$ is the disjoint set formed by $L \cap R$ and
$Dj.L.1$ is the disjoint set formed by $L \backslash (L \cap R)$.
Each of these disjoint sets will be page-aligned by the
custom linker script.

The pseudocode for creating these disjoint sets
is shown in Algorithm \ref{alg:create-disjoint-sets}.
The algorithm begins with the ``deck sets.''
A deck set corresponds
directly to 1 of the 4 types of decks:
the function of a \textbf{single} deck forms a singleton;
the functions of a \textbf{loop} deck form their own set;
any encompassed function that can be
called from some non-loop path has itself and any \textbf{reachable} functions
as part of a set; and any functions that have their addresses taken
and can be invoked by some \textbf{indirect} call form a set with
their statically reachable callees.
The algorithm begins with a list of these sets
and then iterates, attempting to separate functions into their
own disjoint sets, if possible.
To find the disjoint sets,
each pair of the decks is intersected with each other.
If the intersection is non-null, those shared members
are removed from the pair of decks and form their own disjoint set.
This pairwise intersection-removal process is
repeated until no more disjoint sets can be formed.
Once the disjoint sets are known, each one is
assigned its own page-aligned section in the linker script.

\begin{algorithm}[htb]
\caption{Pseudocode for creating disjoint sets
for \shortname's custom linker script.}
\label{alg:create-disjoint-sets}
\begin{algorithmic}[]
\Function{create\_disjoint\_sets}{$deck\_sets$}
    \State $disjoint\_sets \gets \emptyset$
    \While {$!deck\_sets.\Call{empty}$}
        \State $A \gets deck\_sets[0]$
        \State $tmp \gets deck\_sets[1:]$
        \State $deck\_sets \gets \emptyset$
        \For{$B$ in $tmp$}
            \State $I \gets A \cap B$
            \State $A_{I} \gets A \setminus I$
            \State $B_{I} \gets B \setminus I$
            \If{$|I| == 0$}
                \State $deck\_sets.\Call{push}{B}$
                \State \textbf{continue}
            \EndIf
            \If{$|A_{I}| == |B_{I}| == 0$}
                \State \textbf{continue}
            \EndIf
            \If{$|B_{I}| > 0$}
                \State $deck\_sets.\Call{push}{B_{I}}$
            \EndIf
            \State $deck\_sets.\Call{push}{I}$
            \State $A \gets A_{I}$
        \EndFor
        \State $disjoint\_sets.\Call{push}{A}$
    \EndWhile
    \State \Return $disjoint\_sets$
\EndFunction
\end{algorithmic}
\end{algorithm}

\subsection{Runtime component} \label{runtime-component}
The runtime support is exposed as a library to the application.
It is responsible for enabling and disabling pages
by marking them RX or RO.
The API is shown in Listing \ref{lst:runtime-api}. These align directly
with the 4 types of decks mentioned previously (single, loop,
reachable, and indirect), plus library initialization.
Although not shown in the listing,
these API calls also have a corresponding deck teardown
call to remap the relevant pages RO
(and initialization has a corresponding destroy call).

Two important steps are necessary for library initialization.
The first is identifying the binary's base address.
Function offsets are known at build-time, but at runtime
\shortname still needs to determine the text section's
base address. The second step is to protect all of the
text pages by marking them RO. This happens at the start of main,
and main is left RX.

When the program is running, any of the 4 main API endpoints
may be invoked.
\texttt{deck\_single} takes as argument the function ID of an impending
callee. The runtime uses the ID to look up the actual page address
of this function, and it marks it as RX. When that
function returns, \texttt{deck\_single\_end} will mark the page as RO.
\texttt{deck\_reachable} is similar. It takes a callee function ID
as its only argument. Because the callee is an encompassed
function, though, all statically reachable functions must be marked
RX, as well. Because these are compile-time known, the runtime library
only needs to issue a map lookup to find which set of functions
to mark RX for that particular callee.

\texttt{deck\_loop} takes a single loop ID parameter as its argument.
A unique loop ID is assigned by the compiler to each interprocedural,
outermost loop in the program. When this runtime endpoint
is invoked, it is a simple lookup to find which functions are part
of that loop, and to map them RX.

\texttt{deck\_indirect} takes a runtime address as its argument. This
is mapped by the library to the corresponding function, in order
to determine whether that function is an encompassed or non-encompassed
function. If it is encompassed,
then the library leverages its own \texttt{deck\_reachable}
support for that function. Similarly, non-encompassed functions
are handled by the \texttt{deck\_single} support.

\shortname maintains a reference counter for the text pages.
When a function is needed, the reference count for each
of its pages is incremented;
when that function is no longer needed, the reference count
for each of its pages is decremented.
Whenever a page's count changes from 0 to 1,
the page must be marked RX;
and whenever a page's count changes from 1 to 0,
it can be marked RO again.
We define the set of pages at runtime with
reference counts greater than 0 as the \textbf{available pages}.
Adding a deck at runtime will either increase
the cardinality of the \textbf{available pages} (if new pages
are needed), or have no effect on its size (if all needed pages
are already available); the opposite holds for removing a deck.




\begin{lstlisting}[
  caption={\shortname's runtime API.},
  label=lst:runtime-api,
  language=C,
  style=c,
  numbers=none
]
int deck_init(void);
int deck_single(int callee_func_id);
int deck_loop(int loop_id);
int deck_reachable(int callee_func_id);
int deck_indirect(long long callee_addr);
\end{lstlisting}

\subsection{Optimizations} \label{optimizations}
We discuss one performance and one security optimization
that we implemented in \shortname.
We call these two optimizations \textbf{indirect deck caching}
(IDC) and \textbf{stack cleaning} (SC), respectively.
Refer to Section \ref{eval-optimizations} for their evaluation.

The most important performance optimization in \shortname is
reducing the overhead of indirect decks inside
of loops. Instrumenting
inside loops and just before indirect calls can reduce
jump targets to a single function, but this comes at the
expense of runtime overhead.
To improve this, a combination of inlining and caching
can eliminate nearly all overhead.

When any loop execution encounters an indirect call at runtime,
the IDC optimization enforces an inlined check against the
function pointer address. If the function pointer address
is already cached by the runtime library, then the
jump target's page(s) must already be RX, and the application
can proceed without any library call overhead.
If not, then the application pays some performance cost
for invoking \texttt{deck\_indirect} on that iteration:
The secure runtime library makes the page(s) available
and caches the function pointer address in a hashmap.
Note that the inlined code in the application is only for
reading from the map, and the cache is cleared on loop exit.
The intuition is that the combination of inlining the check,
where the hardware's branch predictor can be effective,
and using a hashmap, where library
call overhead is eliminated and the system's memory cache can
be leveraged, should severely limit extra cycles inside of loops.

For security, the SC optimization attempts to further reduce
the attack surface. The general idea is to destroy decks 
in the call stack, and to only reconstruct them when returning
up the call chain. This raises a question as to how
much of the call stack should be ``clean,''
i.e. marked RO and unavailable.
For example, SC could ensure only the last 4 decks are available
on the call stack.
Currently \shortname only implements SC for single decks,
and it imposes the strictest depth of 2.
Thus, when \texttt{deck\_single} is invoked,
SC only allows the current function and its upcoming callee to
be mapped. Similarly, when \texttt{deck\_single\_end} is invoked,
SC only allows the current function and its parent to be mapped.

\section{Evaluation}\label{sec:evaluation}
We perform experiments on the SPEC CPU 2017 suite \cite{spec2017},
10 programs from the GNU coreutils package \cite{coreutils},
and the nginx web server v1.20.1 \cite{nginx}.
We perform all experiments on a commodity desktop running
an AMD Ryzen 7 1800X with 32GB RAM on Ubuntu 18.04 LTS.
Our compiler is based on LLVM 11.0.0.
Unless stated otherwise, \shortname's results are with
the IDC and SC optimizations enabled (see Section \ref{optimizations}).
Here we present our performance and security findings.
Our evaluation focuses on the following questions:
\begin{enumerate}
    \item How much slowdown does an application incur when using
    the \shortname framework? What is the code growth due to linker optimization for segregating function sets?
    \item What is the gadget reduction for applications using \shortname?
    \item How effective are the optimizations in terms of performance and security?
    \item In addition to general gadget reduction,
    can \shortname break real gadget chains in the benchmarks and
    real-world applications to be able to stop gadget-based attacks?
\end{enumerate}

\subsection{SPEC CPU 2017}
SPEC CPU 2017 \cite{spec2017} is a staple suite for CPU-bound performance
benchmarking, making it useful for stressing the performance
of binaries running under \shortname.
It also includes a diverse group of applications that
give us insight into \shortname's effect on security, too.
We use C and C++ applications from the suite, which
are used in a variety of domains, including
route planning, discrete event simulation,
video compression, alpha-beta tree search,
molecular dynamics, and ray tracing.

\begin{figure}[h]
    \centering
    \includegraphics[scale=0.35]{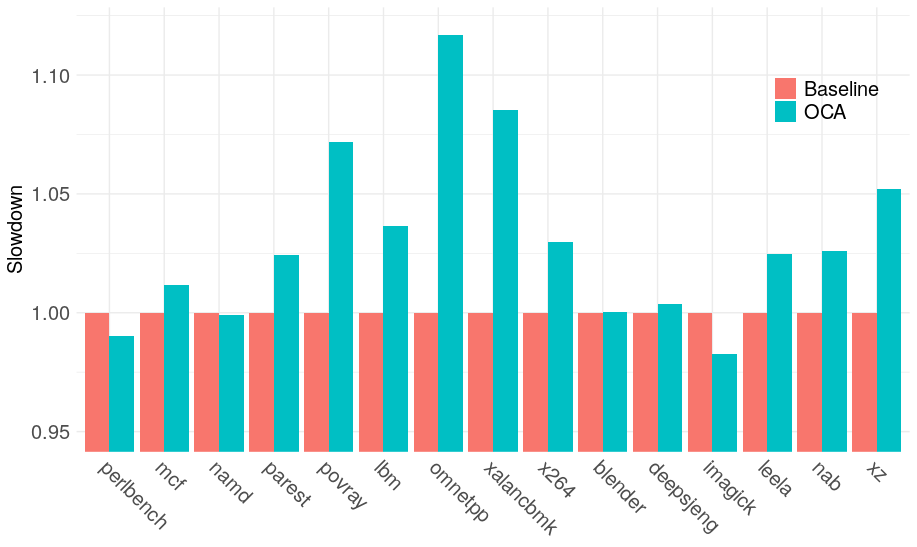}
    \caption{Slowdown for SPEC CPU 2017 using \shortname (normalized against a baseline).}
    \label{figure:spec-perf}
\end{figure}

The performance results are reported in Figure \ref{figure:spec-perf}.
We compile and run a baseline version
of each benchmark, optimized at -O3. Then we recompile and run
the benchmark with \shortname.
The worst-case slowdown is 11\% for \texttt{omnetpp}.
Both \texttt{imagick} and \texttt{perlbench} have slight speedups, which can
happen in instrumentation-based works that affect memory
alignment \cite{cpi, picfi, bincfi, blankit};
\shortname also modifies function layout across pages
which may play a role.
The average slowdown across SPEC is 4\%.

In comparison, BlankIt achieves 18\% overhead
on SPEC CPU 2006 by debloating libraries (not the application).
Razor achieves 1.7\% overhead on average on SPEC CPU 2006,
with a worst-case of 16\%. Piece-wise adds only negligible
load-time overheads but deals only with libraries and not whole applications.
Thus, we find \shortname's 4\% average slowdown (on whole applications) in SPEC CPU 2017
to be reasonable compared with existing approaches.

\begin{table}[h!]
 \centering
  \caption{SPEC CPU 2017 total gadget reduction as a percentage (higher is better).}
 \begin{tabular}{|l|l|l|l|}
   \hline
   \textbf{Application} & \bf Min & \bf Max &  \bf Avg \\
   \hline
   \hline
    perlbench	&	51.1	&	98.8	&	68.4 \\
    mcf	&	25.6	&	68.4	&	54.0 \\
    namd	&	75.4	&	94.0	&	88.5 \\
    parest	&	76.1	&	99.8	&	94.6 \\
    povray	&	36.6	&	97.4	&	53.0 \\
    lbm	&	47.9	&	62.9	&	57.4 \\
    omnetpp	&	52.4	&	98.4	&	79.1 \\
    xalancbmk	&	58.9	&	99.6	&	72.8 \\
    x264	&	17.2	&	99.9	&	32.5 \\
    blender	&	73.9	&	99.8	&	98.5 \\
    deepsjeng	&	24.4	&	68.2	&	64.9 \\
    imagick	&	39.3	&	99.4	&	88.7 \\
    leela	&	54.6	&	87.7	&	84.2 \\
    nab	&	68.6	&	91.9	&	86.6 \\
    xz	&	51.7	&	94.9	&	74.1 \\
    AVERAGE	&	50.2	&	90.7	&	73.2 \\
   \hline
  \end{tabular}
  \label{table:spec-gadget-reduction}
\end{table}

Calculating gadget reduction for \shortname is more complicated than in
pre-runtime techniques like Razor,
where the binary is trimmed, and the gadget
reduction is measured and recorded offline.
In \shortname, however, the gadgets that
are available to an attacker change dynamically during runtime
based on which pages \shortname has mapped RX, i.e. based on the
\textbf{available pages} (discussed in Section \ref{runtime-component}).

To calculate these values, we start with
the total number of gadgets in the application.
This is the sum of ROP, JOP, COP, and special-purpose gadgets
(reported by ROPGadget \cite{ropgadget}):
\begin{equation}
T = R + J + C + S
\end{equation}
The reduction of total gadgets for some set of available
pages $AP$ during runtime is given by the equation:
\begin{equation}
reduction_{AP} = \frac{T - T_{AP}}{T} * 100
\end{equation}
---where $T$ is the total number of gadgets in the baseline
application, and $T_{AP}$ is the total number of gadgets
in that available page set.
Note, to calculate $T_{AP}$, this is equivalent to Equation (1)
for that specific available page set. For example, $R_{AP}$
is calculated as:
\begin{equation}
R_{AP} = \sum_{ap} R_{ap}
\end{equation}
---which is the sum of ROP gadgets $R$ over each available page $ap$ in
the available page set $AP$.
Thus, the average reduction over all available page sets
is equivalent to:
\begin{equation}
avg\_reduction = \frac{\sum reduction_{AP}}{num\_APs}
\end{equation}

To capture these reduction metrics, we first
enable \shortname's logging and run the program under all test inputs.
This dumps the available pages on every \shortname API call to a
log file.
Next, we scan every log line and identify the gadgets
across each available page set.
Each set of available pages
is considered ``equal'' to another for the purposes of gadget-counting.
The set of available pages that has the least proportion of reduced gadgets
is marked as the ``mininum reduction,'' and vice versa
for maximum reduction.
The average reduction is the average proportion of gadgets
reduced by each set of available pages over the logs.


The total gadget reductions are reported
in Table \ref{table:spec-gadget-reduction}.
Each benchmark has
three columns for minimum, maximum, and average reduction in total gadgets.
\shortname achieves an average of 73.2\% total gadget reduction
across all SPEC CPU 2017 benchmarks.
The totals gadget reductions are representative of the individual
(ROP, JOP, COP, and special-purpose) results,
which is expected.
For example, \shortname's average reduction
of ROP gadgets specifically is 77.3\%.
\shortname is not designed for reducing or favoring
any particular type of gadget.
Because it works at function and page granularity, it
has a very similar reduction across all gadget types.

Direct comparisons are difficult because of differences
in technique or reporting.
Piece-wise reduces total gadgets by an average of 72.88\%
on SPEC CPU 2006 benchmarks for musl-libc.
BlankIt reports an average of 97.8\% ROP gadget reduction
on SPEC CPU 2006 benchmarks for all libraries (and using glibc).
Razor reports 68.19\% code reduction (not gadgets) for
applications in SPEC CPU 2006.
Thus, \shortname's SPEC security result appears
to be similar to the other application-focused technique,
Razor, without sacrificing soundness.
Compared with the library-only techniques,
\shortname appears to reduce applications' gadgets equally
as well as the load-time technique (Piece-wise),
but not as thoroughly as an on-demand runtime technique (BlankIt).

\subsection{GNU coreutils}
We measure our technique on a subset of GNU coreutils.
This package contains roughly 100 tools,
including \texttt{grep}, \texttt{mkdir}, and \texttt{rm}.
These utilities are relevant to software debloating for several
reasons, including their real-world ubiquity, and that they
have a history of CVEs. Chisel and Razor also report
results for coreutils that \shortname can compare against.

The authors of Razor made their tool available \cite{razor-github},
so we use the same application versions and inputs as them.
Their inputs were designed to cover the same
functionality tested by Chisel.
We use only the test inputs, as we do not require
any training. The number of inputs per benchmark ranges
between 17-40, and the number of options that any
given input may exercise ranges from 1-7 (see \cite{razor}
for more details).

Runtime overheads are negligible for coreutils.
Every test completes in under 1 second and is trivially performant.
In contrast, SPEC CPU 2017 tests each take 3-10 minutes.

\begin{table}[h!]
 \centering
  \caption{GNU coreutils total gadget reduction as a percentage (higher is better).}
 \begin{tabular}{|l|l|l|l|}
   \hline
   \textbf{Application} & \bf Min & \bf Max &  \bf Avg \\
   \hline
   \hline
    bzip2	&	42.7	&	78.8	&	70.8 \\
    chown	&	88.3	&	97.3	&	95.9 \\
    date	&	95.0	&	97.5	&	96.9 \\
    grep	&	65.0	&	90.9	&	82.8 \\
    gzip	&	34.7	&	75.7	&	64.6 \\
    mkdir	&	90.4	&	96.6	&	94.5 \\
    rm	&	88.4	&	98.7	&	96.9 \\
    sort	&	79.0	&	91.9	&	90.5 \\
    tar	&	49.0	&	86.8	&	83.4 \\
    uniq	&	93.0	&	96.0	&	95.4 \\
    AVERAGE	&	72.5	&	91.0	&	87.2 \\
   \hline
  \end{tabular}
  \label{table:coreutils-gadget-reduction}
\end{table}

As with SPEC, we present the total gadget reduction numbers
(see Table \ref{table:coreutils-gadget-reduction}).
The average decrease across coreutils is 87.2\%.
The worst-case scenario occurs at one point during \texttt{gzip},
where only 34.7\% of the application's gadgets are unavailable.
The best case occurs for \texttt{rm} at 98.7\%.
In comparison, Razor and Chisel
achieve 61.9\% and 85.1\% ROP gadget reduction on coreutils,
respectively.
Thus, \shortname compares well against two other
state-of-the-art techniques that reduce ROP gadgets
in application code.

\subsection{nginx}

\texttt{nginx} is by some metrics the most popular web server today \cite{w3techs}.
Besides serving web content, it is also deployed frequently as
a reverse proxy or load balancer. As a common multitool
in today's web infrastructure, security is a real concern for \texttt{nginx}.
It is multithreaded and multiprocessed, unlike the SPEC benchmarks
and coreutils applications we have evaluated, and this can break or
stress frameworks. \texttt{nginx}'s performance is a critical factor
in certain deployments, so it is important that any security techniques
not interfere too heavily with it.
It is also recently evaluated by another debloating technique, BlankIt,
which will serve as a good comparison point.
For all these reasons, we choose to evaluate \texttt{nginx} with \shortname.

We faithfully reproduce the security and performance experiments
described in the BlankIt evaluation \cite{blankit}.
We use the same \texttt{nginx} workload generator, \texttt{wrk},
which runs 12 threads
in parallel and creates 400 concurrent connections to the server.
For the performance experiment, there are 4 inputs:
the home page of Wikipedia, and 3 randomized binaries of
1MB, 10MB, and 100MB. The performance experiment includes
2 separate tests. In the first test, \texttt{wrk}
requests the Wikipedia home page for 3s, then 30s, then 300s.
The experiment tests that \texttt{nginx} can serve a normal-sized page
(80KB) under high load for extended periods of time without
degrading.
In the second test, \texttt{wrk} requests the 1MB binary for 30s,
then the 10MB binary for 30s, and finally the 100MB binary for 30s.
This experiment tests that as the request size scales,
there is still no degradation.
For the security experiment, only the home page of Wikipedia is used.
\texttt{wrk} makes requests for 30s.

\begin{figure}[h]
    \centering
    \includegraphics[scale=0.35]{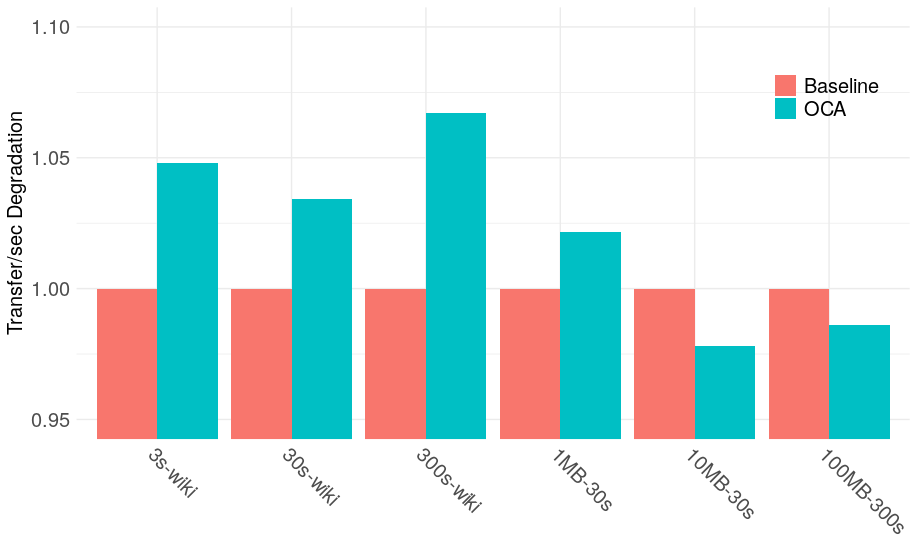}
    \caption{Transfer/sec degradation for \texttt{nginx} using \shortname (normalized against a baseline).}
    \label{figure:nginx-perf}
\end{figure}

The performance result is shown in Figure \ref{figure:nginx-perf}.
The slowdown is reported as the transfer/sec degradation,
normalized against the baseline. \shortname achieves 1.023x slowdown
on average.
Table \ref{table:nginx-gadget-reduction} shows the total
gadget reduction on \texttt{nginx}.
The average is 80.3\% (an improvement over SPEC, but less than
the reduction for coreutils).

\begin{table}[h!]
 \centering
  \caption{nginx total gadget reduction as a percentage (higher is better).}
 \begin{tabular}{|l|l|l|l|}
   \hline
   \textbf{Application} & \bf Min & \bf Max &  \bf Avg \\
   \hline
   \hline
    nginx	&	50.3	&	95.3	&	80.3 \\
   \hline
  \end{tabular}
  \label{table:nginx-gadget-reduction}
\end{table}

In comparison, BlankIt averages 1.047x runtime overhead
and 98.9\% ROP gadget reduction on \texttt{nginx}'s libraries.
As with SPEC, \shortname outperforms BlankIt at runtime but
with less ROP gadget reduction. BlankIt copies
needed library code into place before use and zeroes it out after
use. This accounts for BlankIt's higher gadget reduction, and also
explains why, despite only being used on libraries, BlankIt
is slower than \shortname's page-mapping scheme.

\subsection{Benefits from Optimization} \label{eval-optimizations}
We evaluate two optimizations, indirect deck caching (IDC)
and stack cleaning (SC) (see Section \ref{optimizations}).
We also evaluate the effect of the modified linker script
used by \shortname.

IDC can be broken into two parts: inlining a check inside loops
to see if a function pointer is already mapped; and waiting until
loop exit to disable the function pointer's page(s).
We found that both of these steps are critical to good performance.
For example, SPEC's \texttt{xz} application suffers from over
10x slowdown without IDC. Disabling the page(s)
after loop exit reduces the slowdown to under 8x. Inlining
the check for the mapped pages and using caching reduces the
slowdown to only 1.05x. The inlined check is the major
factor across benchmarks. For example, \texttt{perlbench},
\texttt{x264}, and \texttt{povray} have over 5x, 2.5x,
and 1.3x slowdown, respectively, without it.

In terms of security, IDC and the custom linker script (CLS)
are highly beneficial. \shortname achieves 
28.3\% total gadget reduction on SPEC without either of these
techniques. Applying only CLS, this improves to 43.1\%. Applying
only IDC, it improves to 66.4\%. Together they achieve
the 73.2\% total reduction reported previously.
IDC's benefit to attack surface reduction is a strong argument
against fully static treatment of function pointers, which leaves
too many gadgets available at loop headers.
The SC optimization, however, seems to have little effect.
On SPEC, we measured only negligible improvement in total gadget
reduction (<1\%). Because it is
conservative (being applied only on single decks) and does
not negatively impact performance, we have left it active
in cases where security could benefit.
The result suggests that SC would need to be expanded
in future work to handle other types of decks.

\subsection{Binary Size Growth}
We measured the binary size increase over every application.
There is 2.9x increase across SPEC, 1.5x across
coreutils, and 1.8x for \texttt{nginx}.
In absolute terms, the modified binaries are 17MB,
1.2MB, and 7.1MB on average. Thus, for these applications,
the binaries are still reasonably sized, despite the
growth, and the performance measurements have
confirmed that this has not adversely affected runtime.
The improvement of coreutils over SPEC is due to
fewer disjoint sets in coreutils. All
binaries in coreutils have fewer than 200 sets, whereas
the majority in SPEC have 200 or more.

The worst-case growth without any custom linking would
assign a page-aligned section to every function in the program
(i.e. every disjoint set would be a singleton).
We estimate this case, lower-bounding it
by ignoring weak function symbols in the baseline
applications. Our custom linking script
improves over lower-bounded, worst-case growth by
1.8x, 2.7x, and 1.3x for SPEC, coreutils, and
\texttt{nginx}, respectively.

\subsection{Breaking ROP Gadget Chains}

Gadget reduction is a common metric in security-focused debloating
works \cite{piece-wise,blankit,razor,chisel,trimmer}, but
it is still difficult to draw certain conclusions. For example,
even with 90\% total gadget reduction, a significantly
large enough binary could still have an enormous 
number of gadgets in absolute terms.
Alternatively, 50\% reduction may be
good, provided that it removes a critical class of gadgets
needed by any useful chain.
Thus, gadget reduction is a useful
yardstick, but we also need to investigate whether
an attack surface reduction technique actually
removes attacks.


Ropper \cite{ropper} is an open source tool that can
automatically build gadget chains given some constraints.
It allows us to automatically test if we can successfully build
a ROP gadget chain that spawns \texttt{/bin/sh} via an \texttt{execve}
syscall.
We use Ropper to identify which binaries from all our previous
experiments have this ROP gadget chain. Then we
test every binary over all test inputs with \shortname
and check every available page set for the ROP gadget chain.

Table \ref{table:gadget-chain-breaking} shows the prevalence
of the ROP chain across all applications' text sections.
(\texttt{blender}'s analysis times out, so it is not included.)
Recall from Listing \ref{lst:rop-chain-execve} that
this ROP chain requires a gadget to store values in memory, a sequence
of gadgets to set up the syscall arguments, and a gadget
to invoke the \texttt{execve} syscall.
These correspond to Table \ref{table:gadget-chain-breaking}'s
W-W-W, Args, and Syscall columns, respectively.
(W-W-W stands for a write-what-where gadget.)
If all of these components of the gadget chain are present
for an application, then the end-to-end exploit exists
(a checkmark in the ``E2E exploit'' column).
Out of 25 baseline applications, 8 of them have the full ROP gadget
chain. 12 have an incomplete chain, and 5 have no components
of the chain.


\textbf{\textit{We analyze every available page set over all
inputs across all applications with Ropper and find that
\shortname does not allow the ROP chain under any set of
dynamic decks.}}
The analysis includes 5,390 unique dynamic deck sets
(with 12,718 dynamic execution count) over all
applications and inputs.

We analyze these chains and their broken counterparts under
\shortname to validate the result and understand how \shortname
is achieving 100\% effectiveness at breaking this ROP chain.
We show an analysis of \texttt{sort} to explain how the syscall
opcode manifests and why \shortname removes it in all
test runs. 
The syscall opcode in x86\_64 is 0x0F05.
Ropper identifies a syscall opcode in \texttt{sort}
within a jump instruction. An objdump with 1 line of
context is shown here:

\begin{verbatim}
4276ae:   41 b0 79          mov    $0x79,%r8b
4276b1:   e9 0f 05 00 00    jmpq   427bc5
4276b6:   80 fa 8a          cmp    $0x7b,%dl
\end{verbatim}

The jump target is \texttt{strftime\_case\_.254+0xfb5}.
The \texttt{e9} byte is the jump opcode, and \texttt{0f 05 00 00} is
the addend to a base (which in little endian is 0x0000050F).
The base is calculated from the program counter
(of the next instruction), which is 0x4276b6 in this case.
Thus the jump target is
0x4276B6 + 0x050F, which yields 0x427BC5.
Notice that it is by chance that the jump target offset happens
to be 0x050F.
This code is part of the \texttt{strftime\_case} function.
This is a custom version of \texttt{strftime} from glibc
that is built into coreutils. As
such, it would not have been handled by load-time techniques
like Piece-wise or library-only techniques like BlankIt.
In these experiments, this function's page is never mapped RX
for \texttt{sort}, and thus \shortname eliminates this syscall opcode.

\begin{table}[h!]
 \centering
  \caption{Execve-to-shell ROP chain availability across all baseline applications.}
 \begin{tabular}{|l|l|l|l|l|}
  \hline
  \textbf{App} & \bf W-W-W & \bf Args & \bf Syscall & \bf E2E exploit  \\
  \hline
  \hline
    perlbench	&	\cmk	&	\cmk	&	\cmk	&	\cmk  \\
    mcf	&	 	&	 	&	 	&	  \\
    namd	&	 	&	 	&	\cmk	&	  \\
    parest	&	\cmk	&	\cmk	&	\cmk	&	\cmk  \\
    povray	&	\cmk	&	\cmk	&	\cmk	&	\cmk  \\
    lbm	&	\cmk	&	 	&	 	&	  \\
    omnetpp	&	\cmk	&	\cmk	&	\cmk	&	\cmk  \\
    xalancbmk	&	\cmk	&	 	&	\cmk	&	  \\
    x264	&	\cmk	&	\cmk	&	\cmk	&	\cmk  \\
    deepsjeng	&	 	&	 	&	 	&	  \\
    imagick	&	\cmk	&	\cmk	&	\cmk	&	\cmk  \\
    leela	&	\cmk	&	 	&	 	&	  \\
    nab	&	 	&	 	&	 	&	  \\
    xz	&	\cmk	&	\cmk	&	 	&	  \\
    bzip2	&	 	&	 	&	 	&	  \\
    chown	&	\cmk	&	 	&	 	&	  \\
    date	&	\cmk	&	 	&	\cmk	&	  \\
    grep	&	 	&	 	&	\cmk	&	  \\
    gzip	&	\cmk	&	 	&	 	&	  \\
    mkdir	&	 	&	 	&	 	&	  \\
    rm	&	\cmk	&	 	&	 	&	  \\
    sort	&	\cmk	&	 	&	\cmk	&	  \\
    tar	&	\cmk	&	 	&	 	&	  \\
    uniq	&	\cmk	&	\cmk	&	\cmk	&	\cmk  \\
    nginx	&	\cmk	&	\cmk	&	\cmk	&	\cmk  \\
  \hline
  \end{tabular}
  \label{table:gadget-chain-breaking}
\end{table}

\section{Related Work} \label{sec:related-work}
Researchers have developed multiple defenses
to deal with code reuse attacks.
One of the most prominent techniques is control flow integrity
(CFI) \cite{cfi,xfi}.
CFI limits forward control flow transfers to legal
edges in the control flow graph (CFG) and callgraphs. It is
typically paired with a shadow stack, which restricts backward
control flow transfers to the legal target on the stack.
CFI has a rich history, addressing a variety
of scenarios, contexts, and attacks
\cite{control_flow_locking, kcofi, griffin, uCFI, patharmor, flowguard, opaque_cfi, practical_cfi, typearmor, pittypat, mcfi, cfi-lb}.

CFI defends against ``control data attacks,''
but it does not directly protect against
``non-control data attacks'' \cite{non_control_data},
which corrupt data but not code pointers.
Like control data attacks, they can also be used
to launch code reuse attacks.
Castro et al. introduced data-flow integrity (DFI) in 2006 \cite{dfi},
which helps defend against both control and non-control data attacks.
DFI tracks values and ensures they are not updated
improperly between moves (i.e. that they adhere to the original data
flow of the program).

Memory protection techniques attempt to protect memory itself, i.e. 
prevent the corruption as opposed to just the exploit.
Softbound+CETS provides complete memory safety at 116\%
runtime overhead without hardware support
\cite{softbound, hardbound, cets}.
Code pointer integrity \cite{cpi} provides protection
on only code pointers, which keeps overhead reasonable, but leaves
non-control data attacks out of scope.

Despite these advancements, current state-of-the-art still
has its shortcomings.
There are numerous examples of how to bypass CFI (e.g. 
\cite{control-jujutsu}) or what its limits are
(e.g. \cite{control_flow_bending}). In fact
recent work \cite{cracks} thoroughly categorizes
the shortcomings of several CFI systems, and which they
broadly characterize as:
imprecise analysis methods,
improper runtime assumptions,
unprotected corner code,
unexpected optimization,
incorrect implementation,
mismatched specification,
and unintended targets.
For example, $\pi$CFI \cite{picfi}, which leverages Modular CFI (MCFI)
\cite{mcfi},
uses structural equivalence for type checking. This would
treat \texttt{void *} the same as \texttt{char *}, for instance.
This imprecision allows extra indirect targets for the attacker
to use. $\mu$CFI \cite{uCFI} cannot protect against code pointer reuse and VTable
attacks.
OS-CFI \cite{oscfi} fails to protect against tail calls that are optimized for
indirect calls.

As mentioned, non-control data attacks are out of scope for CFI and CPI.
Though DFI provides some protection for such attacks,
data flow analysis is still
an overapproximation and therefore insecure. Furthermore,
the seminal DFI paper achieved 44-103\% runtime overhead on SPEC CPU 2000 \cite{dfi},
which is generally considered too slow for practical use.
Similarly, Softbound+CETS, despite offering full memory protection,
suffers from these high overheads.

Thus, traditional techniques like CFI for guarding against control flow
hijacking have had
their shortcomings acknowledged. DFI, CPI, and memory protection also
have certain disadvantages that make them in some way incomplete (e.g.
because their defense is incomplete, their scope is limited, or their
overheads are not acceptable for a certain task).
In short, opportunities for attackers to launch code reuse
attacks exist today and are expected to continue.
This has led to recent work that attempts to reduce software's attack surface.

Software debloating and attack surface reduction
are an \textit{orthogonal} solution to approaches such as
CFI or DFI, and can be seen as a \textit{security hardening}
technique.
In addition to \cite{piece-wise, chisel, razor, blankit},
other prominent work includes feature-based techniques.
Slimium \cite{slimium} debloats Chromium features based
on a static-, dynamic-, and heuristic-based analysis.
Koo et al. take a configuration-driven approach to remove
feature-specific code
\cite{config-debloat}, achieving 77\% debloat on \texttt{nginx}.
Trimmer \cite{trimmer} is another technique that takes as input a user
configuration and uses it to drive the debloating process.

Lastly, software engineering
researchers have worked on debloating, but the focus has
not traditionally been on security. For instance,
\cite{causes_of_bloat,making_sense_debloat,analysis_debloat,four_trends_debloat,finding_utility_debloat,container_debloat,finding_reusable_debloat}
leverage debloating to improve performance, and
~\cite{Beszedes:2003:SCR:937503.937504,
Debray:2000:CTC:349214.349233,Muth:2001:ALO:370365.370382,
Franz:1997:SB:265563.265576} use it to reduce code size.


\section{Conclusion}\label{sec:conclusion}
We present \shortname, an attack surface reduction
technique that works on full programs and
can enable may-use code on-demand.
It achieves state-of-the-art
gadget reduction results without compromising
soundness or requiring training or user inputs.
Total gadget reduction across SPEC CPU 2017, GNU coreutils,
and the nginx server average 73.2\%, 87.2\% and 80.3\%,
respectively. In our performance experiments,
the runtime slowdown on SPEC is 4\% and negligible
for GNU coreutils; the transfer/sec degradation for
nginx is only 2\%. In an additional study over
these applications, we show that for all test
inputs, \shortname eliminates the presence of a
ROP chain that spawns a shell via \texttt{execve}.
Based on these results and the generality of the approach,
we find \shortname to be a promising technique for attack
surface reduction.



\bibliographystyle{plain}
\bibliography{refs.bib}

\appendix
\section{Appendix}
This appendix includes: An expanded callgraph of \texttt{xz},
which is helpful for the example in Section \ref{compiler-linking},
and a brief discussion of future work.

\subsection{Expanded \texttt{xz} Callgraph} \label{appendix:expanded-xz-cg}

\begin{figure}[H]
    \centering
    \includegraphics[scale=0.17]{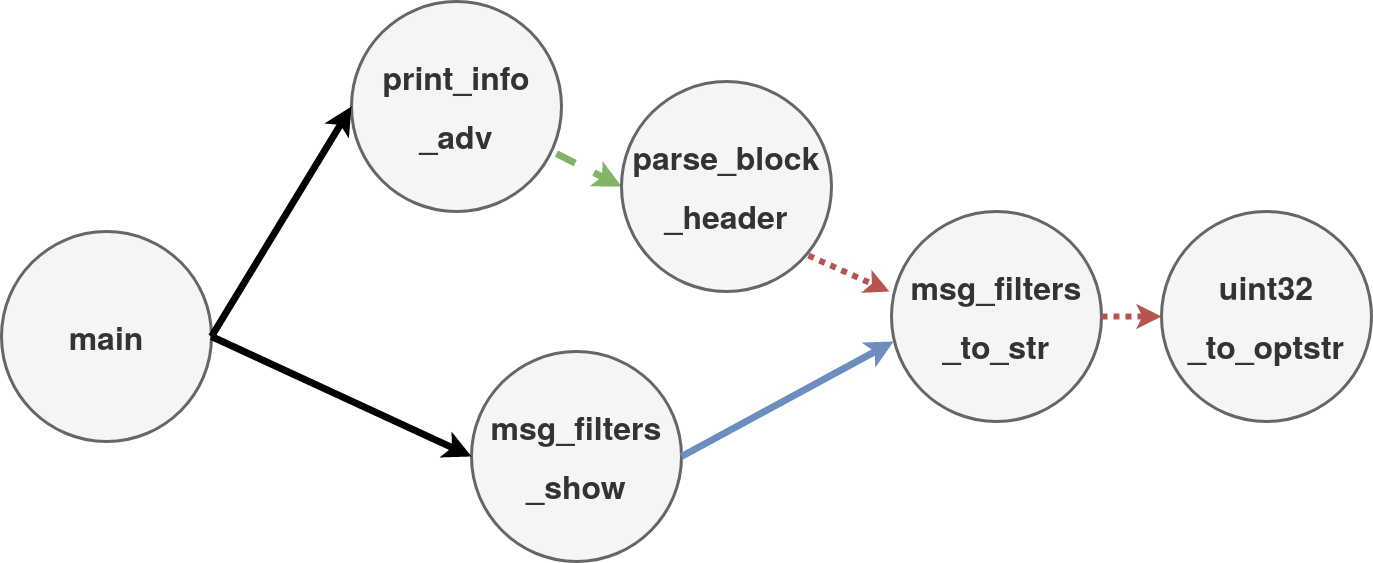}
    \caption{Expanded callgraph from the \texttt{xz} data compression
    application. This expanded graph includes one additional node,
    \texttt{parse\_block\_header}.}
    \label{figure:cg-instrumentation-expanded}
\end{figure}

\subsection{Future Work} \label{appendix:future-work}
There are several interesting directions that we did not explore
with \shortname. The first is leveraging
Intel's MPX \cite{intel-manual} hardware for secure hashing
for the IDC optimization.
In \cite{oscfi}, the authors show how to repurpose
MPX to securely hash and store metadata. Because support
for it is being removed from GCC, it is available for
tools like \shortname to trial.
This could improve IDC's performance further, and in fact,
it opens the possibility of applying
this caching optimization to other types of decks inside of loops.
If a loop exercises only a sparse subset of its
interprocedural, statically reachable function set, then this
could potentially have large security benefits.

We also identified the Intel Memory Protection Keys (MPK) \cite{intel-manual}
hardware primitive as a way to drastically reduce the
runtime's overhead for remapping pages. \shortname's runtime
could potentially replace \texttt{mprotect} calls with
with a single WRPKRU instruction. The libmpk
library \cite{libmpk} makes this secure and simpler, and reports
8.1x speedup for equivalent \texttt{mprotect} calls.
A related insight is that the granularity for building
and tearing down decks does not necessarily need to be at
the function and page level. For that matter, the granularity does
not need to be fixed. By leveraging MPK, for example,
a deck could be composed of a set of basic blocks within a loop,
and good performance might still be achievable.




\end{document}